\documentclass[sigconf]{acmart} 

\usepackage{graphicx}
\usepackage{hyperref}

\usepackage{amsmath,amssymb}
\usepackage{xcolor}
\usepackage{float}

\def \stm{\texttt{StateFi }}

\title{\stm: Effectively Identifying Wi-Fi Devices through State Transitions} 

\author{Abhishek K. Mishra}
\affiliation{%
  \institution{Inria}
  \city{Lyon}
  \country{France}
}
\email{abhishek.mishra@inria.fr}

\author{Mathieu Cunche}
\affiliation{%
  \institution{INSA-Lyon, Inria, CITI Lab.}
  \city{Lyon}
  \country{France}
}
\email{mathieu.cunche@insa-lyon.fr}

\setcopyright{none} 
\acmConference{}{}{}
\acmBooktitle{}
\acmPrice{}
\acmDOI{}
\acmISBN{}

\begin{document}

\begin{abstract}
Randomized MAC addresses aim to prevent passive device tracking, yet Wi-Fi management frames still leak structured behavioral patterns. Prior work has relied primarily on syntactic probe-request features such as Information Elements (IEs), sequence numbers (SEQ), or RSSI correlations, which degrade in dense environments and fail under aggressive randomization. We introduce \stm, a fingerprinting framework that models device behavior as finite-state machines (FSMs), capturing both structural transition patterns and temporal execution logic. These FSMs are embedded into compact feature vectors that support efficient similarity computation and supervised classification. Across five heterogeneous campus environments, \stm achieves \textbf{94--97\%} accuracy for in-network fingerprinting using full management-frame FSMs. With probe-only FSMs, it re-identifies devices under MAC randomization with up to \textbf{97\%} accuracy across large public datasets comprising more than a million frames. When looking at the discrimination accuracy of the model, \stm reaches \textbf{98\%}, outperforming the strongest prior signature by up to \textbf{17 percentage points}. These results demonstrate that FSM-level behavioral dynamics form a powerful and largely unmitigated side channel, stable enough to defeat randomization and expressive enough for robust, scalable device identification.
\end{abstract}


\maketitle

\section{Introduction}

Wi-Fi devices continuously emit unencrypted management frames for discovery, authentication, and association. Although essential for connectivity, these frames leak structured behavioral patterns that enable device fingerprinting and tracking even in the absence of explicit identifiers~\cite{9296300,8936381,9524956}. To mitigate such threats, modern operating systems deploy \emph{MAC address randomization}~\cite{mathieunotenough2016}, frequently rotating locally administered MACs during probe-based discovery. Yet randomization obscures only identifiers, not the underlying behavioral logic encoded in management-frame sequences, timing, and retry dynamics~\cite{tan2021efficient,mishra2024bleach}. Consequently, randomized identifiers provide incomplete protection against behavioral re-identification.

Prior efforts to defeat MAC randomization have relied predominantly on probe-request based signatures derived from Information Elements (IEs), timing cues, sequence numbers (SEQ), or RSSI correlations~\cite{tan2021efficient,mathieunotenough2016,mishra2024bleach}. Although these features can perform well in controlled settings, their discriminative power degrades sharply in realistic environments with high contention, mobility, and heterogeneous AP deployments, and they become increasingly brittle under aggressive randomization~\cite{mishra2023introducing}. Moreover, probe activity is inherently sparse and highly variable, further limiting the stability of such syntactic signatures.

\textbf{In contrast, this paper introduces \stm, which shifts from content-based features to \emph{behavioral dynamics} and analyzes the full spectrum of management frames, not just probe requests.} We model device behavior as a \emph{finite-state machine} (FSM) that captures how management-frame subtypes evolve over time, including both structural transition patterns (e.g., retries, self-transitions) and temporal execution logic (e.g., inter-state delays, scan pacing). These FSMs are then embedded into compact feature vectors that support scalable similarity computation and supervised classification.

Our key insight is that management-plane behavior arises from deep firmware and chipset design choices: scan strategies, association sequences, backoff timing, that remain stable across MAC randomization events and are difficult to obfuscate without harming performance. Through extensive real-world evaluation, we show that these FSM-derived behavioral signatures are both distinctive and persistent.

We evaluate \stm in two operational regimes. In a \emph{network-internal} setting where devices could even remain associated, full-management FSMs achieve \textbf{94--97\%} accuracy across five heterogeneous campus environments, enabling high-precision behavioral integrity monitoring and intrusion detection. In the adversarial \emph{pre-association} setting where devices emit randomized MACs, probe-only FSMs re-identify devices with \textbf{92--97\%} accuracy across three large public datasets containing millions of management frames captured over long durations. Using the \texttt{Infocom'21}~\cite{tan2021efficient} discrimination-accuracy metric, \stm attains \textbf{93--98\%}, outperforming the strongest published signature (IE+SEQ+RSSI) by up to \textbf{17\%}. These gains persist across all sizes of time-window in which we perform MAC de-randomization, demonstrating robustness to contention, mobility, and burst fragmentation.

\medskip
\noindent\textbf{This paper makes the following contributions:}
\begin{itemize}
    \item \textbf{Behavioral Modeling via FSMs:} We introduce the first fingerprinting framework that captures both structural and temporal management-frame dynamics using finite-state machine modeling.
    \item \textbf{Compact and Scalable Embeddings:} We derive an efficient vector representation of FSMs that supports scalable distance-based matching and supervised classification. 
    \item \textbf{Comprehensive Real-World Evaluation:} Across university campus captures and large public datasets, \stm achieves high accuracy for in-network fingerprinting and under MAC randomization, improving on the state-of-the-art by up to \textbf{17\%}.
    \item \textbf{Revealing a Behavioral Side Channel:} We show that MAC randomization fails to conceal implementation-level behavioral fingerprints that could be exploited through FSMs, motivating the need for behavioral obfuscation in future wireless standards.
\end{itemize}

\section{Related Work}

Early work on Wi-Fi device fingerprinting focuses predominantly on extracting \emph{content-based} features from probe requests, with particular emphasis on Information Elements (IEs). Approaches such as~\cite{vanhoef2016wifi, tan2021efficient, mishra2024characterizing, mishra2024fingerprinting, robyns2017noncooperative, pang200711, cunche2014linking} analyze transmission rates, SSIDs, vendor tags, and other IE fields to construct device signatures. Vanhoef et al.~\cite{vanhoef2016wifi} systematically evaluate the discriminability and stability of these fields, proposing strategies for selecting robust IE subsets. Despite their historical prominence, IE-based techniques suffer from well-documented limitations: devices from the same vendor frequently share identical fields, operating system updates alter emitted IEs, and field-level identifiers can be easily obfuscated. Our work departs from these brittle metadata fingerprints by instead capturing persistent behavioral signals expressed through management-frame dynamics.

Beyond content analysis, several studies explore \emph{timing-based} fingerprinting. Clock-skew approaches~\cite{jana2008fast, lanze2012demystified} exploit low-level oscillator imperfections but require precise measurement hardware and extended observation intervals, reducing their practicality in real-world deployments. Other methods~\cite{mishra2021wifi, mathieunotenough2016} extract statistical timing features between frames, yet their performance degrades significantly when probe activity is sparse or when channel contention introduces jitter. In contrast, our framework captures temporal consistency across the \emph{full management lifecycle}: authentication, association, and scanning, yielding fingerprints that are more stable and robust across heterogeneous environments.

Sequence-number–based correlation has also been examined as a means of linking frames emitted by the same device~\cite{vanhoef2016wifi, tan2021efficient}. The state-of-the-art Infocom'21 framework~\cite{tan2021efficient} combines sequence-number continuity, IE fields, and multi-sniffer RSSI observations using a minimum-cost flow formulation to associate randomized MAC addresses with physical devices. While effective in controlled settings, these methods depend on fragile syntactic fields and multi-sensor coordination. Other lines of research circumvent MAC randomization through protocol-level or cryptographic side channels (e.g., UUID-E reversal~\cite{vanhoef2016wifi}), though such attacks rely heavily on specific implementation flaws.

Our approach diverges fundamentally from prior content-, timing-, and sequence-based techniques by modeling \emph{behavioral dynamics} using finite-state machines (FSMs). Rather than just isolating individual frame types or metadata fields, we characterize the complete structure and timing of management-plane interactions. This includes probe exchanges as well as authentication and association behaviors, providing a holistic view of how devices execute the Wi-Fi management protocol in practice.

\section{Data Collection and Preprocessing}\label{data}

Our evaluation uses eight datasets spanning both \emph{in-house campus captures} and \emph{external large-scale public traces}. The campus datasets consist of passive recordings of all management frames across five heterogeneous environments: Café, Laboratory, Classroom, Residence, and Restaurant, chosen to reflect diverse device densities, mobility patterns, and infrastructure setups. To assess robustness under strong MAC randomization, we additionally include three widely used public datasets: \textbf{Infocom'21}~\cite{tan2021efficient}, \textbf{Cagliari}~\cite{pintor2025wifi}, and \textbf{MITIK}~\cite{silva2025mitik}. Together, these datasets enable a comprehensive evaluation under both \emph{network-internal} and \emph{pre-association} threat models.

\subsection{Capture Settings and Dataset Overview}

Campus captures were collected fully passively, ensuring device behavior remained unmodified. Sessions lasted 30–95 minutes and yielded over 2.6M packets. Table~\ref{tab:dataset_summary} summarizes the five internal environments that provide a realistic distribution of contention and mobility.

\begin{table}[ht!]
\centering
\caption{Summary of the Campus dataset}
\label{tab:dataset_summary}
\small
\begin{tabular}{lrrrr}
\toprule
\textbf{Env.} & \textbf{Duration (min)} & \textbf{MACs} & \textbf{Packets} & \textbf{Mgmt.} \\
\midrule
Residence         & 80  & 1,526  & 304k  & 51k  \\
Cafe              & 68  & 1,782  & 80k   & 21k  \\
Laboratory        & 66  & 2,448  & 321k  & 35k  \\
Classroom         & 93  & 4,726  & 541k  & 59k  \\
Restaurant        & 85  & 4,735  & 207k  & 43k  \\
\midrule
\textbf{Total} & -- & 31,862 & 2.6M+ & 235k \\
\bottomrule
\end{tabular}
\end{table}

To evaluate performance under randomized MACs, we also use three external benchmark datasets:

\begin{itemize}
    \item \textbf{Infocom'21}~\cite{tan2021efficient}: high-density probe requests collected via 21 ceiling-mounted sensors on 2.4\,GHz channel~1 across an $\sim$8000\,m$^2$ shopping mall floor. A single business hour contains $\sim$20k frames and $\sim$5k unique MACs.
    \item \textbf{Cagliari}~\cite{pintor2025wifi}: a labelled ``Individual Devices'' subset with probe requests captured on channels 1/6/11 from 18 devices across six operating modes, each in 30-minute sessions.
    \item \textbf{MITIK}~\cite{silva2025mitik}: four 60-minute passive experiments using five supersniffers (25 radios) at La Rochelle University, producing 2.65M probe requests and responses and 5,425 unique MACs.
\end{itemize}

These external traces form the basis of our probe-only evaluations in Section~\ref{subsec:external-macrand}, where the goal is to link randomized MAC addresses across bursts in the pre-association phase.

\subsection{Privacy Considerations and Processing}

All campus traces undergo a unified privacy-preserving sanitization workflow aligned with institutional ethics protocols and local data-protection regulations. We strip all personally identifiable fields: including MAC addresses, SSIDs, and absolute timestamps, retaining only anonymized metadata such as management-frame subtype and inter-arrival timing. Malformed or unrecognized frames are discarded, and only valid management, control, and data frames are preserved. To ensure that our analysis focuses exclusively on client-side behavior, we apply a multi-stage filtering pipeline: 20,615 of the 32,310 observed MACs ($63.8\%$) are flagged as randomized based on local-administration bits and short-lived rotation patterns; and 265 MACs transmitting \texttt{Beacon} or \texttt{Association Response} frames, indicative of access-point behavior, are removed. This processing ensures that all subsequent fingerprinting models operate solely on client-side, implementation-driven behavioral characteristics.

\subsection{Isolating Bursts and Ground Truth}
\label{bursts}

Following the methodology from \texttt{Infocom'21}~\cite{tan2021efficient} and BLEACH~\cite{mishra2024bleach}, we define a \emph{burst} as a sequence of packets emitted by the same MAC address where inter-packet gaps never exceed one second. This models natural scanning activity, where probe requests and related management frames occur in tightly spaced clusters.

Because the true device identity is not observable in any dataset, we rely on persistent MACs as ground truth proxies. For each such MAC, we treat every burst as an independent behavioral instance of the same physical device. Bursts belonging to the same MAC are assigned a unique anonymized identifier, enabling supervised learning and cross-burst consistency evaluation. 

This burst-anchored ground truth is used uniformly across all datasets except \textbf{Cagliari} (where we already have the controlled ground-truth) and evaluation settings—supporting \emph{full-management FSM analysis} for internal fingerprinting (Section~\ref{subsec:campus-envs}) and \emph{probe-only FSM analysis} for randomized MAC re-identification (Section~\ref{subsec:external-macrand}).

\section{FSM-Based Device Characterization}\label{fsm}

Wi-Fi communication follows the IEEE 802.11 state machine~\cite{ieee80211}, which prescribes the logic governing device discovery, authentication, association, and roaming. These interactions unfold through management frames that remain unencrypted even under modern standards (e.g., WPA3) and are therefore observable in passive captures. Crucially, while the standard specifies \emph{allowable} sequences, it does not dictate precise ordering, timing, or retry behavior. Vendors thus implement this logic differently to optimize power consumption, connection latency, roaming responsiveness, or regulatory compliance~\cite{vanhoef2016wifi}. These implementation-level choices introduce subtle but stable variations in management-frame behavior, producing a rich and persistent behavioral footprint. Our framework models these behaviors using \textbf{Finite State Machines (FSMs)} to capture both structural and temporal dynamics that remain stable even under MAC randomization.

\subsection{Modeling Device Behavior as an FSM}

Let $\mathcal{S}$ denote the set of all management-frame subtypes observed across the eight datasets (Section~\ref{data}). For a device $d$, we formalize its behavior as a finite state machine:
\begin{equation*}
\mathcal{F}_d = (Q, \Sigma, \delta, q_0),
\end{equation*}
where:
\begin{itemize}
    \item $Q \subseteq \mathcal{S}$ is the set of states corresponding to management frame subtypes (e.g., \texttt{ProbeReq}, \texttt{Auth}, \texttt{AssocReq}),
    \item $\Sigma$ is the timestamp-ordered sequence of frames emitted during a burst,
    \item $\delta : Q \times Q \rightarrow \mathbb{N}$ is a transition function recording observed transitions between successive subtypes,
    \item $q_0$ is the initial state, set to \texttt{ProbeReq} in probe-only FSMs and to the first observed subtype in full-management FSMs.
\end{itemize}

Transitions are extracted directly from passive traces: a directed edge $q_i \rightarrow q_j$ is recorded whenever frame subtype $j$ immediately follows subtype $i$ for the same device within a burst. Because our model is entirely empirical, capturing what the device \emph{actually} does rather than what the protocol allows, it naturally encodes vendor- and OS-specific logic, retry patterns, scanning aggressiveness, timing behavior, and deviations introduced by energy-saving strategies. 

Empirically, we observe significant structural variation across vendors and device classes. For vendor $v$, let $\mu_v$ denote the average number of transitions per FSM; differences in $\mu_v$ across the dataset reflect heterogeneous scanning intervals, probing frequency, and connection logic. These distinctions persist even when MAC addresses rotate rapidly, supporting the robustness of FSM-based characterization.

\subsection{Constructing FSMs from Bursts}
\label{fsm-burst-construction}

FSMs are built at the granularity of \emph{bursts} (Section~\ref{bursts}), which serve as short, coherent behavioral units. This construction supports both evaluation regimes in the paper:

\begin{itemize}
    \item \textbf{Full-management FSMs (Section~\ref{subsec:campus-envs}):}  
    When MAC addresses remain stable (e.g., after association), we can capture complete management-plane interactions: probe traffic, authentication/association exchanges, reassociation attempts, and deauthentication logic. These FSMs enable high-precision fingerprinting and anomaly detection \emph{within the network}. 

    \item \textbf{Probe-only FSMs (Section~\ref{subsec:external-macrand}):}  
    Under MAC randomization, devices broadcast primarily probe requests and related scanning frames. To remain aligned with the adversarial pre-association setting, we construct FSMs solely from these probe-phase frames (requests and responses). Despite the reduced alphabet, probe-only FSMs retain rich behavioral structure, ordering of IEs, scanning patterns, bursts of identical probes, and chipset-specific retry dynamics. These fingerprints remain stable across randomized MAC addresses, which we will validate in the results section (Section \ref{results}).
\end{itemize}

For each burst, we parse the ordered sequence of management subtypes and create a directed, weighted transition graph. Consecutive occurrences of the same subtype naturally generate self-transitions (e.g., repeated \texttt{ProbeReq} frames), signaling differences in scanning aggressiveness or network-unavailability handling. These per-burst FSMs encode both structural and temporal consistency: in addition to transition frequencies, we record inter-state delays, which reflect implementation-level timing strategies.

Despite sharing the same protocol specification, devices produce structurally distinct per-burst FSMs due to chipset logic, firmware behavior, scheduling policies, and radio coexistence features. These subtle but stable differences form the foundation of StateFi’s ability to (i) fingerprint associated devices with high precision and (ii) re-identify devices across randomized MAC addresses. The burst-level FSM is therefore the core abstraction powering the accuracy and robustness demonstrated in Sections~\ref{results}.

\section{FSM-Based Behavioral Comparison and Fingerprinting}
\label{comparison}

To compare device behavior across heterogeneous environments, \stm models each device's Probe Request sequence as an FSM. However, directly comparing FSM graphs is computationally expensive: graph isomorphism and subgraph matching require combinatorial exploration of node mappings, and differences in state cardinality, transition multiplicity, and burst length further complicate alignment. Such operations scale poorly with the number of devices and the thousands of FSMs produced in realistic traces. To enable large-scale analysis, we instead embed each FSM into a compact numerical representation that preserves its key structural and temporal properties, allowing efficient similarity computation and seamless integration with supervised classifiers.

\subsection{FSM Feature Representation}

Each FSM is embedded into a real-valued feature vector summarizing its state structure, transition dynamics, and temporal behavior. Table~\ref{tab:features} lists the features extracted for every aggregated burst.

\begin{table}[t]
\centering
\caption{FSM-derived features.}
\label{tab:features}
\begin{tabular}{ll}
\toprule
\textbf{Feature} & \textbf{Description} \\
\midrule
$x_1$ & \# unique FSM states (frame subtypes) \\
$x_2$ & \# total state transitions \\
$x_3$ & \# self-transitions ($s_i \!\rightarrow\! s_i$) \\
$x_4$ & Entropy of transition probabilities \\
$x_5$ & Transition rate ($x_2 / \Delta t$) \\
$x_6$ & Mean inter-burst time gap \\
$x_7$ & Max sequence-number gap across bursts \\
$x_8$ & Binary encoding of Probe Request IEs \\
\bottomrule
\end{tabular}
\end{table}

Let $\mathbf{x}_i \in \mathbb{R}^d$ denote the FSM vector for the $i$-th burst group.  
Each vector contains seven FSM-derived features and a binary IE vector of a randomly chosen frame in the burst. This hybrid representation captures both temporal dynamics and protocol-level semantics.








\subsection{Device-Pair Classification}
\label{mls}

To evaluate how reliably we predict that the two bursts originate from the same physical device, \stm adopts a supervised learning pipeline. We construct pairwise similarity features between FSM vectors.

For two bursts with vectors $\mathbf{x}_i$ and $\mathbf{x}_j$, we compute:
\[
\mathbf{f}_{ij} =
\big[
\| \mathbf{x}_i - \mathbf{x}_j \|_2,\;
1 - \cos(\mathbf{x}_i, \mathbf{x}_j),\;
\| \mathbf{x}_i - \mathbf{x}_j \|_1,\;
\text{IE\_sim}(\mathbf{x}_i,\mathbf{x}_j)
\big],
\]
where $\text{IE\_sim}$ is the normalized overlap of the IE bit vectors.  
These features provide complementary views of behavioral similarity: magnitude, direction, structural deviation, and protocol syntax.

\paragraph{Positive and Negative Pair Construction.}
Training data is created by pairing bursts associated with persistent (non-randomized) MAC addresses.  
For each MAC with at least two bursts, we randomly sample up to five \emph{within-MAC} pairs as positives.  
Negative examples are created by pairing bursts from different MACs.  
We balance the dataset by subsampling both categories to a maximum of 5,000 total pairs per scenario.

\paragraph{Model Training and Hyperparameter Selection.}
We evaluate three standard binary classifiers:

\begin{itemize}
    \item \textbf{Random Forest (RF)} with balanced class weights,
    \item \textbf{Support Vector Machine (SVM)} with RBF kernel,
    \item \textbf{Logistic Regression (LR)} with $\ell_2$ regularization.
\end{itemize}

SVM and LR operate on standardized features using scikit-learn's
\texttt{StandardScaler}, while RF operates directly on raw features.
For each scenario and burst-aggregation parameter $P$, we perform
hyperparameter optimization via \texttt{GridSearchCV}. All classifiers
(SVM, Logistic Regression, and Random Forest) and preprocessing tools are
implemented using the \texttt{scikit-learn} machine learning library~\cite{scikit-learn}. 
The final model is trained on 70\% of the constructed pair samples and tested
on the remaining 30\%, and we report accuracy, balanced accuracy, confusion
matrices, and per-class precision/recall. 

\paragraph{Summary.}
This accuracy-focused supervised learning framework enables StateFi to evaluate device-pair classification across heterogeneous datasets while avoiding the prohibitive cost of direct FSM graph comparison. We next present the detailed accuracy results for RF, SVM, and LR.

\section{Results}
\label{results}

We evaluate \stm across complementary settings that reflect the two principal goals of behavioral fingerprinting: (i)~\emph{fingerprinting associated devices and detecting behavioral anomalies within a network}, and (ii)~\emph{re-identifying devices across MAC address randomization during pre-association scanning}. In all cases, accuracy is defined as the proportion of correctly matched communication bursts originating from the same physical device. All methods operate on FSM-based fingerprints constructed from individual bursts. Together, the following subsections demonstrate that StateFi provides both high-precision intrusive-detection capabilities \emph{after} devices join the network and strong re-identification capabilities \emph{before} association, where MAC randomization is intended to protect user privacy. 

\subsection{Accuracy In Fingerprinting devices}
\label{subsec:campus-envs}

We first evaluate \stm in a \emph{network-internal} setting, where the goal is to fingerprint devices and detect anomalous or spoofed behavior even after \emph{after} the association. In this scenario, MAC addresses are relatively stable (per SSID or session), and the objective is not to defeat address randomization, but to determine whether two FSMs correspond to the same associated device and to enable fine-grained behavioral intrusion detection.

To this end, we instrument five heterogeneous campus environments: Café, Laboratory, Classroom, Residence, and Restaurant; and construct FSMs from the \emph{full set of observable management frames} associated with long-lived MAC addresses. This includes probe traffic, but also authentication, association, (re)association, and deauthentication frames exchanged as devices join and maintain connectivity with the campus network. Each FSM thus captures the complete management-plane behavior of a device within a given time window.

\begin{figure}[t]
    \centering
    \includegraphics[width=\linewidth]{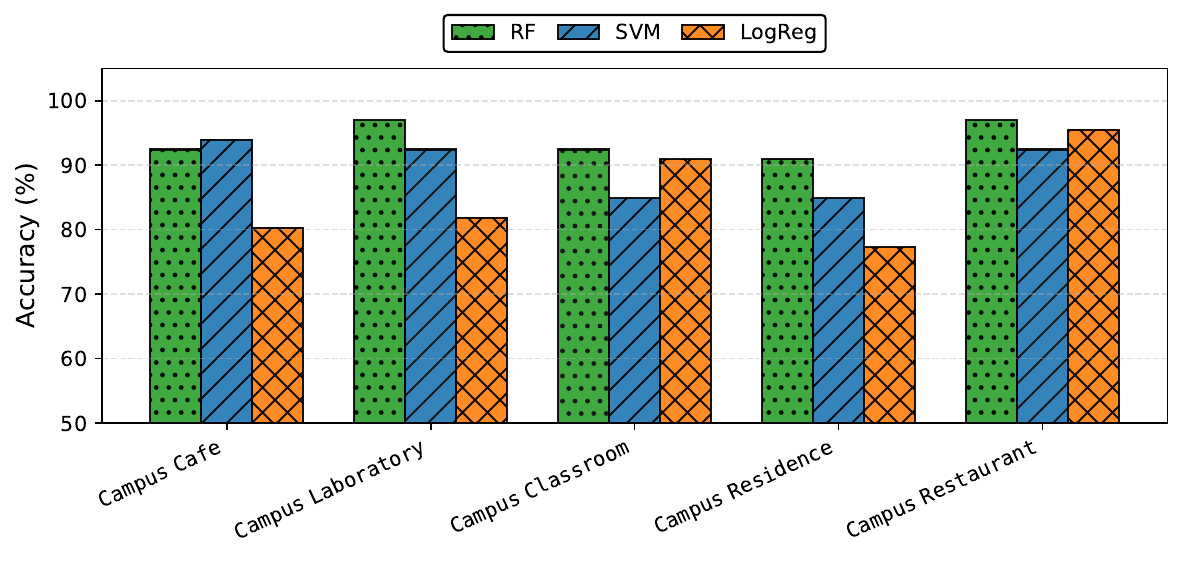}
    \caption{Fingerprinting accuracy across heterogeneous campus scenarios using RF, SVM, and LR models. Here, FSMs are constructed from the full management-frame exchange of associated devices, enabling high-precision behavioral fingerprinting for anomaly and intrusion detection.}
    \label{fig:setA}
\end{figure}

For every scenario, we derive pairwise distance features between FSM embeddings (Section~\ref{mls}) and train three supervised classifiers: Random Forest (RF), Support Vector Machine (SVM), and Logistic Regression (LR). The resulting accuracies are shown in Fig.~\ref{fig:setA}.

Across all environments, RF yields the strongest performance, achieving approximately 94--97\% accuracy, while SVM and LR lag slightly behind. These results demonstrate that full management-frame FSMs provide a stable and highly discriminative behavioral signature once a device has associated with an access point. In other words, even when MAC addresses are only session- or SSID-stable, \stm can reliably distinguish devices and detect deviations in their management behavior, making it well-suited for network-side intrusion detection, device impersonation detection, and integrity monitoring.

\subsection{Accuracy in Defeating MAC Randomization}
\label{subsec:external-macrand}

We next turn to the setting most relevant to MAC address randomization: the \emph{pre-association discovery phase}, where devices broadcast probe requests using frequently changing, locally administered MAC addresses. In this regime, devices are not yet associated with any AP, and the dominant packet type is the Probe Request. To align with this adversarial model, we restrict our FSMs to \emph{probe requests and responses}, and ask whether \stm can still link bursts generated by the same physical device across randomized MACs.

\begin{figure}[t]
    \centering
    \includegraphics[width=\linewidth]{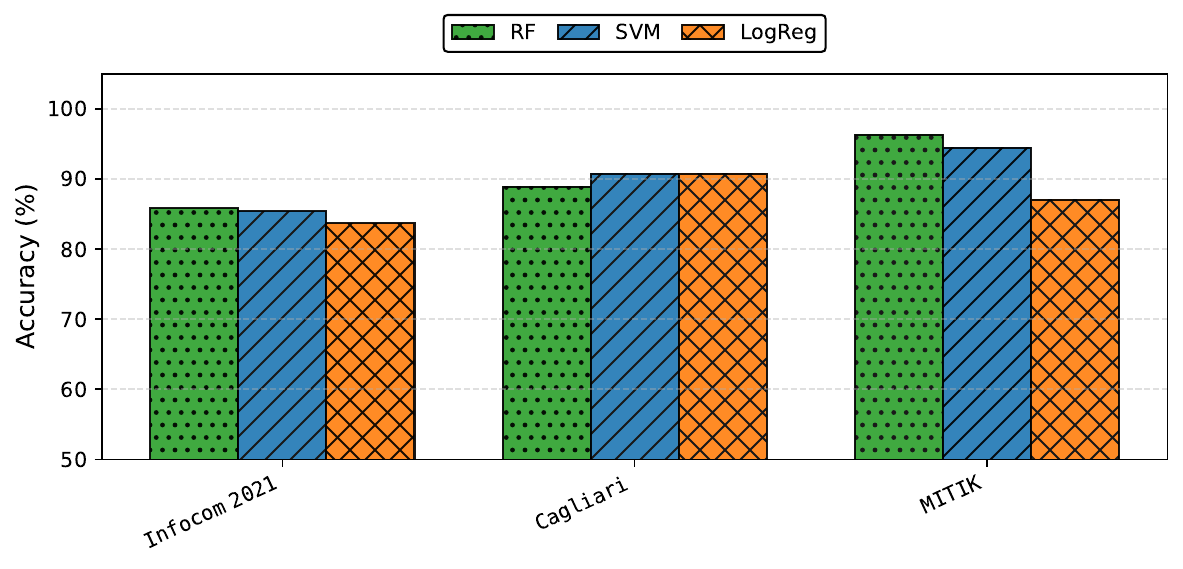}
    \caption{Accuracy in defeating MAC randomization for large public datasets (Infocom'21, Cagliari, MITIK) using probe-only FSMs. \stm maintains high accuracy even when operating solely on pre-association probe behavior, illustrating its ability to re-link devices across randomized MAC addresses.}
    \label{fig:setB}
\end{figure}

We evaluate this capability on three large public datasets: Infocom'21, Cagliari, and MITIK, which capture large numbers of mobile devices in realistic public settings. For each trace, we construct probe-only FSMs by utilizing individual bursts and derive the same pairwise distance features as in the campus experiments. We then train RF, SVM, and LR classifiers to predict whether two probe-only FSMs originate from the same underlying device, despite potential changes in the source MAC address. The resulting accuracies are reported in Fig.~\ref{fig:setB}.

The trends mirror those observed in the campus setting: RF consistently achieves the highest accuracy (typically 92--97\%), with LR following closely and SVM underperforming both. Crucially, these results are obtained using \emph{only} probe request and response behavior, without relying on stable association phases or long-lived MACs. This shows that the FSM of probe activity, i.e., how often, in what patterns, and with which information elements a device scans the environment, is sufficiently distinctive to re-identify devices across randomized MAC identifiers.


\subsection{Comparison with State-of-the-Art}
\label{subsec:infocom-comparison}

To contextualize the performance of \stm within the broader literature on MAC address association, we compare our results with the \texttt{Infocom'21} framework~\cite{tan2021efficient}, which represents the most comprehensive and highest-performing approach reported to date. \texttt{Infocom'21} integrates \emph{all major feature families} used in prior work, including Information Elements (IEs), Sequence Numbers (SEQ), and multi-sniffer RSSI correlations, into a unified association strategy. As shown in~\cite{tan2021efficient}, this combination outperforms techniques relying solely on timing, IEs, sequence numbers, or RSSI individually, and is widely recognized as the state-of-the-art for probe-to-device association under long-lived or session-stable MAC addresses. For these reasons, \texttt{Infocom'21} serves as the strongest available baseline for evaluating whether behavior-based FSM fingerprints provide measurable improvements.

\begin{figure}[t]
    \centering
    \includegraphics[width=\linewidth]{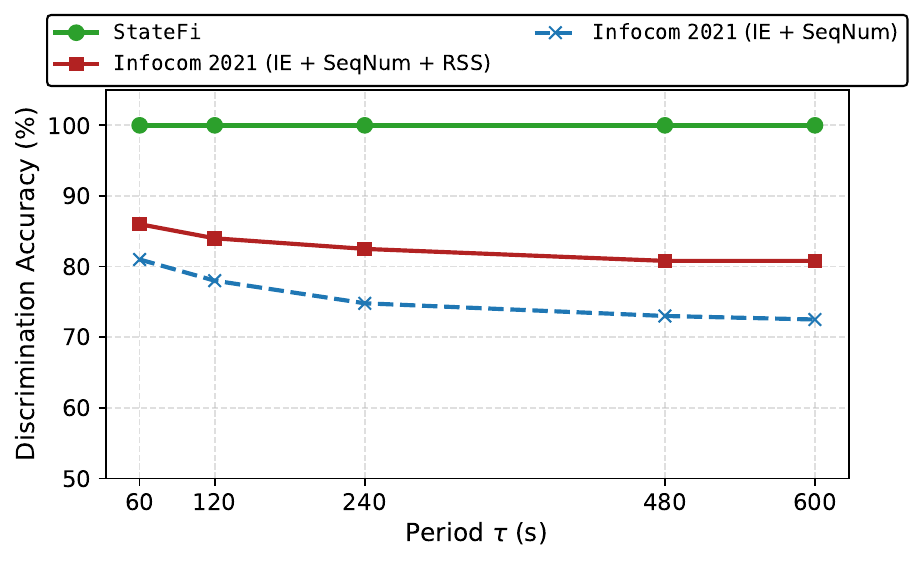}
    \caption{Discrimination accuracy comparison on the Infocom'21 dataset for $\tau \in \{60,120,240,480,600\}$ seconds. \stm substantially outperforms IE- and SEQ-based baselines, including the full IE+SEQ+RSS signature.}
    \label{fig:infocomCompare}
\end{figure}

To enable a faithful comparison, we adopt the \emph{discrimination accuracy} metric introduced in \texttt{Infocom'21}. Given a probe request $p_i$ at time $t_i$, the metric evaluates whether a method can correctly associate $p_i$ to a previous probe from the same physical device within the temporal window $[t_i - \tau, t_i]$. Using the same dataset utilized by \cite{tan2021efficient}, we report the accuracies for their IE+SEQ+RSS signature and the IE+SEQ signature, and compare them to our FSM fingerprint. Figure~\ref{fig:infocomCompare} presents the side-by-side results across standard window sizes $\tau \in \{60, 120, 240, 480, 600\}$ seconds.

StateFi delivers a substantial improvement over both Infocom'21 signatures. Across all time windows, our probe-only FSM fingerprint achieves \textbf{93--98\% discrimination accuracy}, demonstrating both high precision and temporal stability. In comparison, the strongest Infocom'21 baseline (IE+SEQ+RSS) plateaus at \textbf{86\%}, while IE+SEQ achieves at most \textbf{81\%}, with both degrading as $\tau$ increases. This widening performance gap at longer windows reflects the sensitivity of IE and SEQ fields to burst boundaries, channel contention, and scanning dynamics; factors that introduce drift and fragmentation in syntactic signatures.


Overall, this comparison highlights a key advantage of StateFi as FSM-based behavioral modeling provides a stable, implementation-driven signature that retains discriminative power under realistic, noisy, and adversarial operating conditions.

\section{Discussion}


\vspace{0.15cm}
\noindent\textbf{Behavioral Basis of FSM Fingerprints.}
Unlike approaches that rely heavily on volatile syntactic fields such as MAC addresses, IEs, or sequence counters, FSM-based modeling leverages deeper behavioral signals: the order in which devices traverse protocol states and the timing characteristics of those transitions. These patterns arise from firmware logic, chipset timing, scanning policies, and retransmission strategies, all of which remain relatively stable across environments and sessions. By encoding both transition structure and inter-state timing into unified feature vectors, FSMs capture robust behavioral traits that persist even when identifiers are frequently randomized.

\vspace{0.15cm}
\noindent\textbf{Limitations and Countermeasures.}
FSM fingerprints could be weakened by protocol-level obfuscation (e.g., randomized scanning sequences, injected timing noise, or homogenized state-transition profiles) or by reducing visibility into management frames through emerging WPA3 extensions. However, these defenses impose real costs: excessive randomness may slow association, impair roaming, increase power consumption, or conflict with vendor-specific optimizations that improve performance. The practical feasibility of masking behavioral signatures without degrading user experience remains uncertain.

\section{Conclusion}

We introduced \emph{StateFi}, a finite state machine (FSM) based fingerprinting framework that captures structural and temporal management frame behavior to derive stable device signatures, even under strong MAC randomization. Unlike prior syntactic approaches relying on brittle fields such as IEs or sequence numbers, \stm leverages implementation-driven behavioral transitions that are significantly harder to mask or randomize. Across diverse real-world environments, full-management FSMs achieve \textbf{94–97\%} in-network identification accuracy, while probe-only FSMs attain \textbf{92–97\%} accuracy under randomized MACs. \stm further reaches \textbf{93–98\%} discrimination accuracy, outperforming the strongest existing signature by up to \textbf{12–17\%} without requiring multi-sniffer deployments. These results position FSM-level behavioral modeling as a robust and scalable foundation for Wi-Fi fingerprinting, while underscoring the persistence of behavioral side channels and the need for future privacy defenses that reduce FSM-level leakage without degrading network performance.




\bibliographystyle{ACM-Reference-Format}
\bibliography{references}


\end{document}